\documentclass{ifacconf}

\usepackage{amsmath}
\usepackage{amssymb}
\usepackage{graphicx}
\usepackage{color}
\usepackage{xcolor}
\usepackage{dashrule}
\usepackage{cancel}
\usepackage{natbib}

\begin{document}
\begin{frontmatter}

\title{Secure Mode Distinguishability for Switching Systems Subject to Sparse Attacks \thanksref{footnoteinfo}}

\thanks[footnoteinfo]{NOTICE: this is the author’s version of a work that was accepted for publication in IFAC
	2017 World Congress. Changes resulting from the publishing
	process, such as peer review, editing, corrections, structural formatting, and other quality control mechanisms may not be reflected in this document. Changes may have been made to this work since it was submitted for publication. A definitive version was subsequently published in IFAC World Congress, July 2017 (see \citet{GEM_IFAC}).}

\author[First]{Gabriella Fiore}
\author[First]{Elena De Santis}
\author[First]{Maria Domenica Di Benedetto}

\address[First]{Center of Excellence DEWS, Department of Information Engineering, Computer Science and Mathematics, University of L'Aquila, Italy. (E-mail: gabriella.fiore@graduate.univaq.it; elena.desantis,mariadomenica.dibenedetto@univaq.it).}

\begin{abstract}
Switching systems are an important mathematical formalism when dealing with Cyber-Physical Systems (CPSs). In this paper we provide conditions for the exact reconstruction of the initial discrete state of a switching system, when only the continuous output is measurable, and the discrete output signal is not available.
In particular, assuming that the continuous input and output signals may be corrupted by additive malicious attacks, we provide conditions for the secure mode distinguishability for linear switching systems. 
As illustrative example, we consider the hybrid model of a DC/DC boost converter.
\end{abstract}


\end{frontmatter}

\section{Introduction}
In Cyber-Physical Systems (CPSs) physical processes, computational resources and communication capabilities are tightly interconnected. Traditionally, the physical components of a CPS are described by means of differential or difference equations, while the cyber components are modeled by means of discrete dynamics.
Therefore, hybrid systems, that are heterogeneous dynamical systems characterized by the interaction of continuous and discrete dynamics, are a powerful modeling framework to deal with CPSs. Switching systems are an important subclass of hybrid systems and can be viewed as a higher level abstraction of a hybrid model. In general, for switching systems the observability property is concerned with the possibility of reconstructing the hybrid state, i.e. both the discrete state (also called mode or location) and the continuous one.


The focus of this paper is on the exact reconstruction of the discrete state of a switching system, when only the continuous output is accessible and the discrete output is not available, using the formalism introduced in \citet{DeSantis2011807} and developed in \citet{01ftsc}, where the authors
address the problem of discrete state reconstruction for ideal autonomous and controlled switching systems. In this paper, we extend those results, by investigating the scenario in which the continuous input and output signals may be corrupted by additive malicious attacks.
This is motivated by the great importance of security issues for CPSs, where the presence of a feedback loop between the physical processes and the controllers through the communication network may increase the vulnerability of the system to failures or malicious attacks. In this case, security measures protecting only the computational and communication layers are not sufficient for guaranteeing the safe operation of the entire system against the presence of malicious attackers. Therefore, new strategies that explicitly address the strong interconnection between the physical, the computational and the network layers are needed.

There exists a vast literature dealing with security for CPSs (see \citet{amin2009safe} and \citet{7011011} to name a few).
In general, cyber attacks can be divided in two main categories (see \citet{amin2009safe}): deception attacks compromising the integrity of the information, and denial-of-service attacks compromising the availability of the information. 
Recent results focus on the case when the attack is not represented by a specific model, but it is assumed to be unbounded and influencing only a small subset of sensors and/or actuators, i.e., the attack is sparse but its intensity can be unbounded (see \citet{Tabuada6727407}, \citet{Hespanha7171098}, \citet{ShoukryNonlinear}, \citet{shoukry2013event}, \citet{Tomlin:15_2},   \citet{7479119ASV}, \citet{fiore2017secure}).
In \citet{Tabuada6727407} the authors propose a method to estimate the state of a linear time invariant system when an unknown (but fixed in time) set of sensors and actuators is corrupted by sparse deception attacks. They prove that, if the number of corrupted devices is smaller than a certain threshold, then it is possible to exactly recover the internal state of the system, by means of an algorithm derived from compressed sensing techniques. A more computational efficient version of the algorithm is presented in \citet{shoukry2013event} where the authors introduce a notion of strong observability and a recursive algorithm that estimates the state despite the presence of the attack.
The same assumption on the sparsity of the attack signal is made in \citet{chang2015secure} where the authors consider a more general case in which the set of attacked nodes can change over time.
A similar approach is used in \citet{Hespanha7171098} where a continuous time linear system is considered.
In order to overcome the limitations imposed by the combinatorial nature of the problem, in \citet{7479119ASV} the authors formulate the problem as a satisfiability one, and propose a sound and complete algorithm based on the Satisfiability Modulo Theory paradigm.
All the above-mentioned works are concerned with the state estimation for linear systems and cannot be directly applied to switching systems.

\subsubsection{Contribution.} In this paper, we investigate under which conditions the exact reconstruction of the initial discrete state of a switching system is resilient against the presence of a sparse attack.
In order to estimate the current mode of the switching system when the discrete output is not available and only the continuous output signal is accessible, we need to \textit{distinguish} which discrete state is indeed active, i.e. we need to \textit{distinguish} between any two dynamical systems, based on the continuous information only.
This problem has been addressed in the literature
for switching systems where the continuous input and output information is not corrupted by failures or malicious attacks (see \citet{01ftsc} and references therein for a complete review of existing results on this topic). In \citet{Baglietto201469} the authors investigate the problem of identifying the current location of a switching system when the continuous measurement signal is corrupted by noise. This disturbance is assumed to have bounded magnitude and therefore their results do not apply to the case in which the additive signal is not a measurement noise but an intentional attack performed by a malicious attacker, the magnitude of which can be unbounded.
Thus, when the continuous input and output signals can be compromised by a malicious attacker, we extend the characterization presented in  \citet{Tabuada6727407} by defining a notion of \textit{secure distinguishability}. We model an attack on the sensors as an attack on the continuous output signal of the switching system, and an attack on the actuators as an attack on the continuous input signal. We consider both the case of autonomous switching systems and the case of controlled switching systems.

The paper is organized as follows.
In Section \ref{sec:ProbForm} we provide a general formulation of our problem. In Section \ref{sec:Controlled} we investigate the case of controlled switching systems in which both the sensor measurements and the actuator signals can be corrupted by a malicious attacker. In Section \ref{sec:Autonomous} we consider the case of autonomous switching systems with sparse attacks on sensors. 
In Section \ref{sec:NumResult} we provide an illustrative example, in which we check if the dynamics of a DC/DC boost converter are securely distinguishable, making use of the hybrid model provided in \citet{Sanfelice7070893}.

\subsubsection{Notation.} In this paper we use the following notation. 
$I$ indicates the identity matrix, $\mathbf{0}$ indicates the null matrix of proper dimensions (which can be trivially deduced by the context).
Given a vector $x \in \mathbb{R}^n$, $\mathrm{supp}(x)$ is its support, that is the set of indexes of the non-zero elements of $x$; $\|x\|_0$ is the cardinality of $\mathrm{supp}(x)$, that is the number of non-zero elements of $x$. The vector $x \in \mathbb{R}^n$	is said to be $s$-sparse if $\|x\|_0\leq s$. 
$\mathbb{S}^n_s$ indicates the set containing all the $s$-sparse vectors $x_i\in \mathbb{R}^n$ such that $\|x_i\|_0\leq s$.
Given the function $y:\mathbb{N}\rightarrow \mathbb{R}^p$,
$y|_{[t_0,t_0+T-1]}$ is the collection of $T$ samples of $y$, i.e. $y|_{[t_0,t_0+T-1]}=(y(t_0)^\top \; y(t_0+1)^\top \;  \cdots \;  y(t_0+T-1)^\top)^\top$.
The function $y$ is said to be cyclic $s$-sparse if, given a set $\Gamma\subset \{1,\dots,p\}$, such that $|\Gamma|=s$, $y(t) \in \mathbb{S}^p_s$ and  $\mathrm{supp}(y(t))\subseteq \Gamma$, for all $t\in \mathbb{N}$.
$\mathbb{CS}^{pT}_s$ is the set containing all the cyclic $s$-sparse vectors $y\in \mathbb{R}^{pT}$.
Given a matrix $M \in \mathbb{R}^{n\times m}$ and a set $\Gamma \subseteq \{1, \dots n\}$, we denote by $\overline{M}_{\Gamma} \in \mathbb{R}^{(n-|\Gamma|) \times m}$ the matrix obtained from $M$ by removing the rows whose indexes are contained in $\Gamma$. If $\Gamma$ is the support of a vector $x \in \mathbb{R}^n$, its complement is $\overline{\Gamma}=\{1, \dots n\}\setminus \Gamma$. Thus, $\overline{M}_{\overline{\Gamma}} \in \mathbb{R}^{|\Gamma| \times m}$ is the matrix obtained from $M$ by removing the rows whose indexes are contained in $\overline{\Gamma}$, or, equivalently, the rows whose indexes are not contained in $\Gamma$.
Given the set $ \Pi \subseteq \{1, \dots m\} $ we denote by $\widetilde{M}_{\Pi} \in \mathbb{R}^{n \times(m-|\Pi |) }$ the matrix obtained from $M$ by removing the columns whose indexes are contained in $\Pi$. 
$\widetilde{M}_{\overline{\Pi}} \in \mathbb{R}^{n \times |\Pi | }$ is the matrix obtained from $M$ by removing the columns whose indexes are not contained in $\Pi$.

\section{Problem Formulation}\label{sec:ProbForm}

In this paper we consider a nominal switching system, where the finite set of discrete states is $Q=\{1,\dots,N\}$.
A linear discrete-time dynamical system $S_i$ is associated to each discrete state $i \in Q$, which is fully described by the tuple $(A_i,B_i,C_i)$ as follows:
\begin{equation}
\begin{aligned}
x(t+1)& =A_{i}x(t) + B_{i}u(t),\;  & x(t_0) &=x_{0} \\ 
y(t)& =C_{i}x(t)
\end{aligned}\label{eq:NominalSwitchingSystem}
\end{equation}
where $t\in \mathbb{Z}$, $\mathbb{Z}$ denotes the set of nonnegative integer numbers, $x(t)\in \mathbb{R}^n$ is the (continuous) state of the system, $y(t) \in \mathbb{R}^p$ is the output measured by the sensors, $u(t) \in \mathbb{R}^m$ is the input sent by the controller to the actuators.
The collection of all subsystems $S_i$, $i \in Q$ is denoted by $\mathcal{S}$. 
We assume that only the continuous input and output signals are known, whereas the initial discrete state $q_0 \in Q$, and the initial continuous state $x_0 \in \mathbb{R}^n$ are unknown.
The switching signal specifies which dynamical system $S_i \in \mathcal{S}$ is currently active in each time instant, that is, which is the current discrete state. 
In this paper we assume that the switching signal is unknown and arbitrary, therefore we do not exploit any information about the underlying graph topology, representing the admissible transitions between discrete states. 
Let $t_0=0$ be the initial time. For a given $T>0$, we assume that no switching occurs in the interval $[0,T]$.
In other words, we assume that the switching system dwells enough time in each discrete mode before a new transition takes place. More specifically, we assume the existence of a minimum dwell time $T$ such that each discrete state remains active for at least $T$ steps.

\begin{figure}
	\centering
	\includegraphics[width=0.6\linewidth]{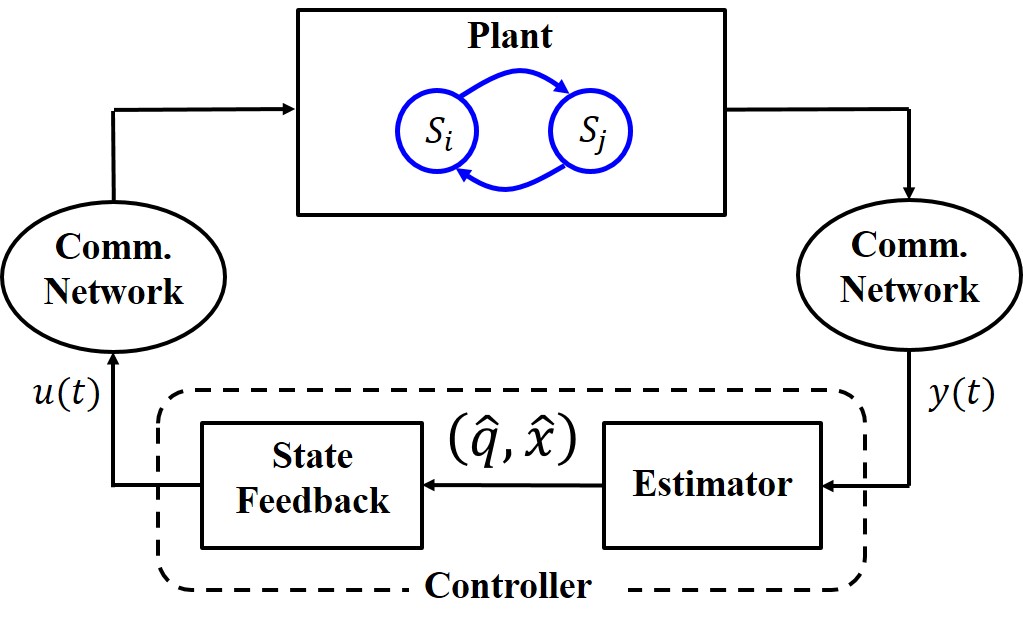}
	\vspace*{-0.3cm}
	\caption[Conceptual block diagram of a wireless control system.]{Conceptual block diagram of the control system. {\footnotesize The plant is modeled as a switching system, a dynamical system as in (\ref{eq:NominalSwitchingSystem}) is associated to each discrete mode $i,j\in Q$. Sensor measurements and control inputs are exchanged by means of a communication network. $\hat{q} \in Q $ indicates the estimate of the current discrete mode, $\hat{x} \in \mathbb{R}^n $ is the estimate of the continuous state.}}
	\label{fig:systemvjpg}
\end{figure}

We consider the scenario in which sensor measurements (continuous output) are sent to the controller, which estimates the true discrete state of the system and the corresponding initial continuous state (which are unknown). Based on this estimation, the controller sends the control signal (continuous input) to the actuators, as shown in Fig. \ref{fig:systemvjpg}. 
We assume that sensor measurements and actuator inputs are exchanged by means of a wireless communication network, and that both of them could be compromised by an external malicious attacker. With this assumption, we consider both the case in which the attacker compromises the devices (sensors or actuators, also called nodes) and the case in which the attacker affects the communication links between different devices (that is, between sensors/actuators and the controller).

The corrupted system can be described as follows:
\begin{equation}
\begin{aligned}
x(t+1)& =A_{i}x(t) + B_{i}[u(t) + v(t)] \\
y(t)& =C_{i}x(t) + w(t)
\end{aligned}\label{eq:CorruptedSwitchingSystem}
\end{equation}
where $v(t) \in \mathbb{S}^m_{\rho}$ is the $\rho$-sparse attack vector on actuator signals, and $w(t) \in \mathbb{S}^p_{\sigma}$ is the $\sigma$-sparse attack vector on sensor measurements.
We assume that the malicious attacker has only access to a subset of sensors $K_w\subset \{1,\dots,p\}$, and to a subset of actuators $K_v\subset \{1,\dots,m\}$, meaning that the set of attacked nodes is fixed over time (but unknown). 
This assumption is motivated by the fact that it is reasonable to suppose that, in a real system, the attacker has not access to the whole set of monitoring and controlling devices. 

\begin{assum} 
	We assume that the attack on sensor measurements is cyclic $\sigma$-sparse (for brevity, $\sigma$-sparse), and that the attack on actuator signals is cyclic $\rho$-sparse (for brevity,  $\rho$-sparse). \label{ass:sparsity}
\end{assum}
\vspace*{-0.2cm}
Roughly speaking, Assumption \ref{ass:sparsity} means that we know that both the set of attacked sensors and the set of attacked actuators have bounded cardinality (that is, $|K_w|\leq \sigma < p$, $|K_v|\leq \rho < m$, respectively), but we do not know which nodes are actually compromised. 
Let $w_k(t)$ denote the $k$-th component of $w(t) \in \mathbb{R}^p$, $k\in \{1,\dots,p\}$ (i.e., the component of $w(t)$ corresponding to the $k$-th sensor), at time $t\in \mathbb{N}$. If $k\notin K_w$, then $w_k(t)=0$ for all $t\in \mathbb{N}$ and the $k$-th sensor is said to be secure (i.e. not attacked). If $k\in K_w$, then $w_k(t)$ can assume any value and this corresponds to the case in which the attacker has access to the $k$-th sensor. The same holds for attacks on the actuators.

The problem that we consider in this paper is to provide conditions for the exact reconstruction of the discrete state of a switching system in the time interval $[0,T]$, based on the knowledge of the corrupted continuous output signal and the continuous input signal (which can be corrupted by a malicious attacker, too). The reconstruction of the discrete mode of the switching system corresponds to understanding which continuous dynamical system is evolving, in a set $\mathcal{S}$ of known ones. This means that, given a pair of linear systems, 
we have to investigate the possibility of \textit{distinguishing} which one of the two systems is active, based on the continuous output and input information, despite the presence of sparse attacks. The initial discrete state can be reconstructed if and only if each pair in $\mathcal{S}$ can be distinguished. If the initial discrete state can be reconstructed, then also the discrete state after each switching can be reconstructed, provided that the dwell time is sufficiently large.

For the nominal system in (\ref{eq:NominalSwitchingSystem}), different distinguishability notions have been proposed, based on the role of the input function and of the continuous initial state (see \citet{DeSantis2011807} for an exhaustive analysis). 
In this paper, we integrate these notions, by introducing the \textit{secure distinguishability} property for the corrupted switching system in (\ref{eq:CorruptedSwitchingSystem}), and we investigate under which conditions this property holds.
More specifically, we assume that the continuous input and output signals can be corrupted by sparse attacks, as in \citet{Tabuada6727407}. However, in spite of considering a discrete-time linear system, we extend the characterization in \citet{Tabuada6727407} to switching systems. Therefore, our attention is focused in providing conditions which enable the correct identification of the current location of the switching system, despite the presence of sparse attacks.

For the sake of clarity, we first review the notion of distinguishability between nominal linear systems, described as in (\ref{eq:NominalSwitchingSystem}), in which the distinguishability 
is required for generic inputs and for all initial states.
A generic input sequence $u|_{[0,\tau-1)}$ is any input sequence that belongs to a dense subset of the set $\mathbb{R}^{m\tau}$, equipped with the $L_\infty$ norm.
Let $y_i$, $i \in Q$, be the output evolution when the dynamical system $S_i$ is active
 with initial state $x(0)=x_{0i}$, and let $\mathcal{U}$ be the set of all input functions $u:\mathbb{N}\rightarrow \mathbb{R}^m$.
\begin{defn}
	(\citet{DeSantis2011807}) Two linear systems $S_i$ and $S_j$, $(i,j)\in Q \times Q$, are input-generic distinguishable if there exists $\tau \in \mathbb{N}$ such that, for any pair of initial states $x_{0i}$ and $x_{0j}$, and for a generic input sequence $u|_{[0,\tau-1)}$, with $u \in \mathcal{U}$, $y_i|_{[0,\tau-1]} \neq y_j|_{[0,\tau-1]}$. The systems $S_i$ and $S_j$ are called indistinguishable if they are not input-generic distinguishable.
	\label{def:IGdist}
\end{defn}


In Section \ref{sec:Controlled} we introduce the notion of secure distinguishability for corrupted discrete time linear systems described as in (\ref{eq:CorruptedSwitchingSystem}), whereas in Section \ref{sec:Autonomous} we describe how to change the notion of secure distinguishability in the case in which autonomous linear systems are considered.

\section{Controlled switching systems}\label{sec:Controlled}

In this section we consider the corrupted system in (\ref{eq:CorruptedSwitchingSystem}), with attacks on sensor measurements and on input signals which are $\sigma$-sparse and $\rho$-sparse, respectively.

First, we recall the result on input-generic distinguishability for the nominal  system in (\ref{eq:NominalSwitchingSystem}). The distinguishability notion implies the comparison between the output evolutions of different dynamical systems. Thus, let two nominal linear systems $S_i$ and $S_j$, $(i,j)\in Q \times Q$, be given. We consider the augmented linear system $S_{ij}$, which is fully described by the triple $(A_{ij},B_{ij},C_{ij})$, such that:
\begin{equation}
\begin{aligned}
A_{ij}=\begin{bmatrix}
A_i & \mathbf{0} \\
\mathbf{0}   & A_j
\end{bmatrix}, \; \; B_{ij}=\begin{bmatrix}
B_i\\ B_j
\end{bmatrix}, \; \; C_{ij}=\begin{bmatrix}
C_i & -C_j
\end{bmatrix}\label{eq:IGmatrix}
\end{aligned}
\end{equation}
with $A_{ij}\in \mathbb{R}^{2n \times 2n}$, $B_{ij}\in \mathbb{R}^{2n \times m}$, $C_{ij}\in \mathbb{R}^{p \times 2n}$.

The following matrices are also associated with the augmented system $S_{ij}$:
\begin{equation}
\begin{aligned}
O_{ij}& ={\footnotesize \begin{bmatrix}
C_{ij} & \\
C_{ij} A_{ij} & \\
\vdots \\
C_{ij} A_{ij}^{2n-1}
\end{bmatrix}= \begin{bmatrix}
O_i^{(2n-1)} & -O_j^{(2n-1)}
\end{bmatrix}} \; , \\
M_{ij}& ={\footnotesize \begin{bmatrix}
\mathbf{0} & \mathbf{0} & \cdots & \mathbf{0} \\
C_{ij} B_{ij} & \mathbf{0} & \cdots & \mathbf{0} \\
\vdots & \vdots & \cdots & \vdots \\
C_{ij} A_{ij}^{2n-2} B_{ij} & C_{ij} A_{ij}^{2n-3} B_{ij} & \cdots & C_{ij} B_{ij}
\end{bmatrix}}
\end{aligned}
\end{equation}

where $\mathbf{0} \in \mathbb{R}^{p\times m}$ is the null matrix, $M_{ij}\in \mathbb{R}^{2np \times (2n-1)m}$.
$O_{ij} \in \mathbb{R}^{2np \times 2n}$ is the $2n$ steps-observability matrix for the augmented system $S_{ij}$, and it is made up of the $2n$ steps-observability matrices $O_{i}^{(2n-1)}$ and $O_{j}^{(2n-1)}$ for the linear systems $S_i$ and $S_j$, respectively.  
\begin{thm}
	(\citet{DeSantis2011807}) Two nominal linear systems $S_i$ and $S_j$, $(i,j)\in Q \times Q$, are input-generic distinguishable if and only if 
$M_{ij}\neq \mathbf{0}$.
	\label{thm:IGdist}
\end{thm}

\begin{assum}
	In this section, we assume that, for any pair $(i,j)\in Q \times Q$, the linear systems $S_i$ and $S_j$ are input-generic distinguishable.
\end{assum}

\vspace*{-0.2cm}
We now extend Definition \ref{def:IGdist} and Theorem \ref{thm:IGdist} to take into account the presence of the (unknown) attack on sensors and actuators, as in (\ref{eq:CorruptedSwitchingSystem}), when the controller is not aware neither of which actuators are corrupted, nor of which sensors are corrupted. In this case, the distinguishability between different modes is required for generic inputs, generic $\rho-$sparse attacks on actuators, for all $\sigma-$sparse attacks on sensors, and for all initial states.

\begin{defn}
	Two linear systems $S_i$ and $S_j$, $(i,j)\in Q \times Q$, are securely distinguishable with respect to generic inputs, generic  $\rho-$sparse attacks on actuators and for all $\sigma-$sparse attacks on sensors (shortly, $\sigma\rho-$securely distinguishable), if there exists $\tau \in \mathbb{N}$ such that $y_i|_{[0,\tau-1]} \neq y_j|_{[0,\tau-1]}$, for any pair of initial states $x_{0i}$ and $x_{0j}$, for any pair of $\sigma-$sparse attack vectors $w_i|_{[0,\tau-1]}\in \mathbb{CS}_{\sigma}^{\tau p}$ and  $w_j|_{[0,\tau-1]}\in \mathbb{CS}_{\sigma}^{\tau p}$, and for any generic  $(u,v_i,v_j)\in \mathcal{U} \times \mathbb{S}_{\rho}^{m} \times \mathbb{S}_{\rho}^{m}$.
	\label{def:CIGAGdistNOACK}
\end{defn}

\begin{figure}
	\centering
	\includegraphics[width=0.7\linewidth]{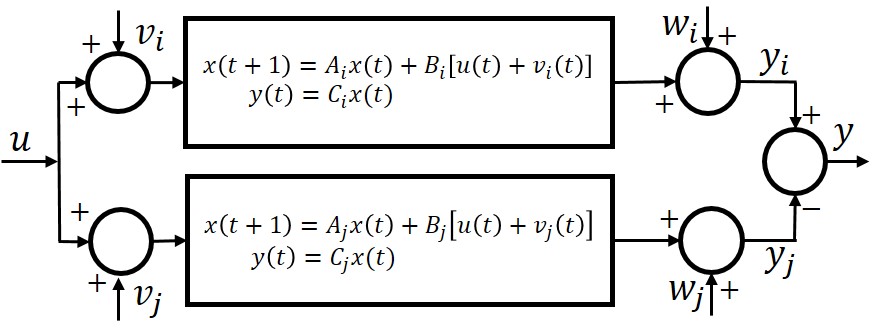}
	\caption[Attack on sensors and actuators.]{Attack on sensors and actuators: the augmented linear system $S_{ij}$.}
	\label{fig:actsensnoackjpg}
\end{figure}

To the aim of characterizing the $\sigma \rho-$secure distinguishability property, we consider the augmented linear system $S_{ij}$ depicted in Fig. \ref{fig:actsensnoackjpg}.
We assume that the sets of attacked actuators for $S_i$ and $S_j$ are, respectively, $\Delta_i\subset \{1,\dots,m\}$ and $\Delta_j\subset \{1,\dots,m\}$, with $|\Delta_i|=|\Delta_j|=\rho$. The sets of attacked sensors for $S_i$ and $S_j$ are, respectively, $\Gamma_i\subset \{1,\dots,p\}$ and $\Gamma_j\subset \{1,\dots,p\}$, with $|\Gamma_i|=|\Gamma_j|=\sigma$.

We can reformulate the component of the state due to the attack on actuators as $B_iv_i=\widetilde{B}_{i,\overline{\Delta}_i}d_i(t)$ and $B_jv_j=\widetilde{B}_{j,\overline{\Delta}_j}d_j(t)$, where  $d_i(t) \in \mathbb{R}^{\rho}$, $d_j(t) \in \mathbb{R}^{\rho}$. $\widetilde{B}_{i,\overline{\Delta}_i} \in \mathbb{R}^{n\times \rho}$, $\widetilde{B}_{j,\overline{\Delta}_j} \in \mathbb{R}^{n\times \rho}$ are the matrices obtained from $B_i$ and $B_j$ by removing the columns whose indexes are not contained in $\Delta_i$ and $\Delta_j$, respectively.

The augmented linear system $S_{ij}$ is represented by the following equations:
\begin{equation}
\begin{aligned}
{\footnotesize x(t+1)} & = {\footnotesize \begin{bmatrix}
A_i & \mathbf{0} \\
\mathbf{0} & A_j
\end{bmatrix} x(t) + \begin{bmatrix}
B_i \\
B_j
\end{bmatrix} u(t) +\begin{bmatrix}
\widetilde{B}_{i,\overline{\Delta}_i} & \mathbf{0}\\ \mathbf{0} & \widetilde{B}_{j,\overline{\Delta}_j}
\end{bmatrix} \begin{bmatrix}
d_i(t) \\ d_j(t)
\end{bmatrix}} \\
& = A_{ij}x(t)+B_{ij}u(t)+\widehat{B}_id_i(t)+\widehat{B}_jd_j(t)\\
y(t) & = \begin{bmatrix}
C_i & -C_j
\end{bmatrix}x(t) + w_i(t) - w_j(t)
\label{eq:IGstateEquationNOack}
\end{aligned}
\end{equation}
where $x(t)\in \mathbb{R}^{2n}$, $w_i(t)\in \mathbb{S}_{\sigma}^p$, $w_j(t)\in \mathbb{S}_{\sigma}^p$, $\widehat{B}_i \in \mathbb{R}^{2n \times \rho}$, $\widehat{B}_j \in \mathbb{R}^{2n \times \rho}$. 

The following matrices are associated with the augmented system $S_{ij}$ in (\ref{eq:IGstateEquationNOack}):
\begin{equation}
\begin{aligned}
O_{ij}& ={\footnotesize \begin{bmatrix}
O_i^{(2n-1)} & -O_j^{(2n-1)}
\end{bmatrix}}  \\
M^{ij}_U& ={\footnotesize \begin{bmatrix}
\mathbf{0} & \mathbf{0} & \cdots & \mathbf{0} \\
C_{ij} B_{ij} & \mathbf{0} & \cdots & \mathbf{0} \\
\vdots & \vdots & \cdots & \vdots \\
C_{ij} A_{ij}^{2n-2} B_{ij} & C_{ij} A_{ij}^{2n-3} B_{ij} & \cdots & C_{ij} B_{ij}
\end{bmatrix}} \\
M_i& ={\footnotesize \begin{bmatrix}
\mathbf{0} & \mathbf{0} & \cdots & \mathbf{0} \\
C_{ij} \widehat{B}_i & \mathbf{0} & \cdots & \mathbf{0} \\
\vdots & \vdots & \cdots & \vdots \\
C_{ij} A_{ij}^{2n-2} \widehat{B}_i & C_{ij} A_{ij}^{2n-3} \widehat{B}_i & \cdots & C_{ij}\widehat{B}_i
\end{bmatrix}} \\
M_j& ={\footnotesize \begin{bmatrix}
\mathbf{0} & \mathbf{0} & \cdots & \mathbf{0} \\
C_{ij} \widehat{B}_j & \mathbf{0} & \cdots & \mathbf{0} \\
\vdots & \vdots & \cdots & \vdots \\
C_{ij} A_{ij}^{2n-2} \widehat{B}_j & C_{ij} A_{ij}^{2n-3} \widehat{B}_j & \cdots & C_{ij}\widehat{B}_j
\end{bmatrix}}
\end{aligned}\label{eq:IFAC:IGmatrices}
\end{equation}
where  $M^{ij}_U\in \mathbb{R}^{2np \times (2n-1)m}$, $M_i\in \mathbb{R}^{2np \times (2n-1)\rho}$, $M_j\in \mathbb{R}^{2np \times (2n-1)\rho}$,
$O_{ij} \in \mathbb{R}^{2np \times 2n}$ is the $2n$ steps-observability matrix for the augmented system $S_{ij}$.

\begin{defn}
	Let two linear systems $S_i$ and $S_j$, $(i,j)\in Q \times Q$, and sets $\Gamma\subset \{1,\dots, p\}$, $|\Gamma|\leq 2\sigma$, $\Delta_k \subset \{1,\dots, m\}$, $|\Delta_k|\leq \rho$, $k=i,j$ be given. Consider the input sequence $u|_{[0,\tau-1)}$, with $u \in \mathcal{U}$, the attack sequences on actuators $d_i|_{[0,\tau-1)}$, $d_j|_{[0,\tau-1)}$ with $d_i(t) \in \mathbb{R}^{\rho}$, $d_j(t) \in \mathbb{R}^{\rho}$, and the attack sequences on sensors $w_i|_{[0,\tau-1)}$, $w_j|_{[0,\tau-1)}$.
	The set $\Omega \subset \mathrm{ker}(\overline{C}_{ij,\Gamma})\subset \mathbb{R}^{2n}$, is defined as:
	\begin{equation}
	\begin{aligned}
	\Omega= & \{ \begin{bmatrix}
	x_{0i}^\top & x_{0j}^\top
	\end{bmatrix}^\top \in \mathbb{R}^{2n} \, : \, \forall u|_{[0,\tau-1)},  \, \forall d_i|_{[0,\tau-1)}, \\  & \forall w_i|_{[0,\tau-1)},  \, \forall w_j|_{[0,\tau-1)}, \, \exists  d_j|_{[0,\tau-1)} : \\
	& y_i|_{[0,\tau-1]} = y_j|_{[0,\tau-1]} \}
	\end{aligned}
	\end{equation}
\end{defn}

Given sets $\Gamma\subset \{1,\dots, p\}$, $\Delta_k \subset \{1,\dots, m\}$ such that $|\Gamma|\leq 2\sigma$, $|\Delta_k|\leq \rho$, $k=i,j$, let $\Omega^{\ast}$ be the maximal subspace such that
\begin{equation}
\begin{aligned}
A_{ij} \Omega^{\ast} + \mathrm{Im}\left(\begin{bmatrix}
B_{ij} & \widehat{B}_i
\end{bmatrix}\right) & \subset \Omega^{\ast} + \mathrm{Im}(\widehat{B}_j) \\
\Omega^{\ast} \subset \mathrm{ker}(\overline{C}_{ij,\Gamma})
\label{eq:OmegaNOack}
\end{aligned}
\end{equation}

Let $\mathcal{W}_{ij,\Gamma}$ be the maximal $\left(  A,
\begin{bmatrix}
B_{ij} & \widehat{B}_i & \widehat{B}_j
\end{bmatrix}
\right)$-controlled invariant subspace contained in $\ker\left(  \overline{C}_{ij,\Gamma}\right)$ (as defined in \cite{basile1992controlled}).

\begin{prop}\label{prop:IG-AG-OmegaNOack}
	Given sets $\Gamma\subset \{1,\dots, p\}$, $\Delta_k \subset \{1,\dots, m\}$ such that $|\Gamma|\leq 2\sigma$, $|\Delta_k|\leq \rho$, $k=i,j$, there exists $\Omega^{\ast}$ as in  (\ref{eq:OmegaNOack})	if and only if:
	\begin{equation}
	\begin{aligned}
	\mathrm{Im}\left(
	\begin{bmatrix}
	B_{ij} & \widehat{B}_i
	\end{bmatrix}
	\right)  \subset \mathcal{W}_{ij,\Gamma}.
	\end{aligned}
	\end{equation}
	If this inclusion holds, then $\Omega^{\ast}=\mathcal{W}_{ij,\Gamma}$.	
\end{prop}

\begin{pf}
	Assume that  $\Omega^{\ast}$ exists.
	Then $\Omega^{\ast}$ is a \\ $\left(  A,
	\begin{bmatrix}
	B_{ij} & \widehat{B}_i & \widehat{B}_j
	\end{bmatrix}
	\right)$-controlled invariant subspace contained in $\ker\left( \overline{C}_{ij,\Gamma}\right)  $,
	hence $\Omega^{\ast} \subset \mathcal{W}_{ij,\Gamma}$ because of maximality of $\mathcal{W}_{ij,\Gamma}$.
	Condition (\ref{eq:OmegaNOack}) implies that $\mathrm{Im}\left(
	\begin{bmatrix}
	B_{ij} & \widehat{B}_i
	\end{bmatrix}
	\right) \subset \Omega^{\ast} \subset \mathcal{W}_{ij,\Gamma}$. On the other side, suppose that
	$\mathrm{Im}\left(
	\begin{bmatrix}
	B_{ij} & \widehat{B}_i
	\end{bmatrix}
	\right) \subset \mathcal{W}_{ij,\Gamma}$. 
	Then $A_{ij}\mathcal{W}_{ij,\Gamma}\subset \mathcal{W}_{ij,\Gamma}+\operatorname{Im}\left(
	\begin{bmatrix}
	B_{ij} & \widehat{B}_i & \widehat{B}_j
	\end{bmatrix}
	\right)
	=\mathcal{W}_{ij,\Gamma}+\operatorname{Im}\left( \widehat{B}_j
	\right)  $.
	Since
	\begin{equation*}
	\begin{aligned}
	A_{ij}\mathcal{W}_{ij,\Gamma}+\operatorname{Im}\left(
	\begin{bmatrix}
	B_{ij} & \widehat{B}_i 
	\end{bmatrix}
	\right)  \subset \mathcal{W}_{ij,\Gamma}+\operatorname{Im}\left(
	\begin{bmatrix}
	B_{ij} & \widehat{B}_i & \widehat{B}_j
	\end{bmatrix}
	\right) 
	\end{aligned}
	\end{equation*}
	and
	\begin{equation*}
	\begin{aligned}
	\mathcal{W}_{ij,\Gamma}+\operatorname{Im}\left(
	\begin{bmatrix}
	B_{ij} & \widehat{B}_i & \widehat{B}_j
	\end{bmatrix}
	\right) =\mathcal{W}_{ij,\Gamma}+\operatorname{Im}\left(\widehat{B}_j\right)
	\end{aligned}
	\end{equation*}
	then $\mathcal{W}_{ij,\Gamma}\subset\Omega^{\ast}
	$, because of maximality of $\Omega^{\ast}$.
	Therefore $\Omega^{\ast}=\mathcal{W}_{ij,\Gamma}$.
	\hfill $\square$
\end{pf}

The output sequence of the augmented system $S_{ij}$ in \eqref{eq:IGstateEquationNOack} can be written in compact form as:
\begin{equation}
\begin{aligned}
Y=O_{ij}^{(\tau-1)}x_0+M_U^{ij}U+M_iD_i-M_jD_j+W_i-W_j
\end{aligned}\label{eq:IFAC:IGcompactNOack}
\end{equation}
where $Y=y_i|_{[0,\tau-1]} - y_j|_{[0,\tau-1]}$, $U=u|_{[0,\tau-1)}$, $D_i=d_i|_{[0,\tau-1)}$,  $D_j=d_j|_{[0,\tau-1)}$, $W_i=w_i|_{[0,\tau-1)}$, $W_j=w_j|_{[0,\tau-1)}$, and the matrices are defined in \eqref{eq:IFAC:IGmatrices}.

\begin{prop}\label{prop:IG-AG-SolNOack} Given sets $\Gamma$, with $\Gamma\subset \{1,\dots , p\}$, $|\Gamma|\leq 2\sigma$ and  $(\Delta_i, \Delta_j)$ with $\Delta_k \subset \{1,\dots,m\}$, $|\Delta_k|\leq \rho$, for $k=i,j$
	\eqref{eq:IFAC:IGcompactNOack} has a solution $\begin{bmatrix}
	x_0^\top & D_j^\top
	\end{bmatrix}^\top \in \mathbb{R}^{(2n+\tau \rho)}$, for all $D_i$, for all $W_i$, for all $W_j$, and for all $U$, if and only if  $\operatorname{Im}\left( 
	B_{ij}
	\right) + \operatorname{Im}\left( 
	\widehat{B}_i
	\right) \subset \mathcal{W}_{ij,\Gamma} + \operatorname{Im}\left( 
	\widehat{B}_j
	\right)$.
\end{prop}

\begin{pf}
	Assume that $y_i|_{[0,\tau-1]} = y_j|_{[0,\tau-1]}$. Then the set $\Omega^{\ast}$ is non-empty. Therefore $A_{ij} \Omega^{\ast} + \mathrm{Im}\left(\begin{bmatrix}
	B_{ij} & \widehat{B}_i
	\end{bmatrix}\right)  \subset \Omega^{\ast} + \mathrm{Im}(\widehat{B}_j)$, and the following holds:		$A_{ij} \Omega^{\ast} + \mathrm{Im}\left(
	B_{ij} \right) + \operatorname{Im}\left( 
	\widehat{B}_i
	\right) \subset \Omega^{\ast} + \mathrm{Im}(\widehat{B}_j) \subset \mathcal{W}_{ij} + \mathrm{Im}(\widehat{B}_j)$.
	Thus $\operatorname{Im}\left( 
	B_{ij}
	\right) +  \operatorname{Im}\left( 
	\widehat{B}_i
	\right) \subset \mathcal{W}_{ij,\Gamma} + \operatorname{Im}\left( 
	\widehat{B}_j
	\right)$. 
		
	Assume now that $\operatorname{Im}\left( 
	B_{ij}
	\right) +  \operatorname{Im}\left( 
	\widehat{B}_i
	\right) \subset \mathcal{W}_{ij,\Gamma} + \operatorname{Im}\left( 
	\widehat{B}_j
	\right)$. Therefore $\mathrm{Im}\left(\begin{bmatrix}
	B_{ij} & \widehat{B}_i
	\end{bmatrix}\right)  \subset \mathcal{W}_{ij,\Gamma} + \mathrm{Im}(\widehat{B}_j)$.
	As $\mathcal{W}_{ij,\Gamma}$ is the maximal $\left(  A,
	\begin{bmatrix}
	B_{ij} & \widehat{B}_i & \widehat{B}_j
	\end{bmatrix}
	\right)$-controlled invariant subspace contained in $\ker\left(  \overline{C}_{ij,\Gamma}\right)$, then there exist $W_i$ and $W_j$ such that $y_i|_{[0,\tau-1]} = y_j|_{[0,\tau-1]}$.
		\hfill $\square$
\end{pf}

\begin{thm}\label{thm:IFAC:ActuatorSensor}
	Two linear systems $S_i$ and $S_j$, $(i,j)\in Q \times Q$, are $\sigma\rho-$securely distinguishable if and only if the following conditions hold:
	\begin{equation}
	\begin{aligned}
	\operatorname{Im}\left( 
	B_{ij}
	\right) + \operatorname{Im}\left( 
	\widehat{B}_i
	\right) \nsubseteq \mathcal{W}_{ij,\Gamma} + \operatorname{Im}\left( 
	\widehat{B}_j
	\right) \\
	\operatorname{Im}\left( 
	B_{ij}
	\right) + \operatorname{Im}\left( 
	\widehat{B}_j
	\right) \nsubseteq \mathcal{W}_{ij,\Gamma} + \operatorname{Im}\left( 
	\widehat{B}_i
	\right)
	\end{aligned}
	\end{equation} 
	for any tuple of sets $(\Gamma,\Delta_i,\Delta_j)$, with $\Gamma\subseteq \{1,\dots , \gamma\}$, $|\Gamma|\leq 2\sigma$, and  $\Delta_k \subseteq \{1,\dots,\delta\}$, $|\Delta_k|\leq \rho$, for $k=i,j$.
\end{thm}

\begin{pf}
	The proof directly follows from Propositions \ref{prop:IG-AG-OmegaNOack} and \ref{prop:IG-AG-SolNOack}.
		\hfill $\square$
\end{pf}

\section{Autonomous switching systems}\label{sec:Autonomous}

\noindent In this section, an autonomous switching system, whose continuous output is corrupted by a sparse attack, is considered. A dynamical system is associated to each discrete state as follows:
\begin{equation}
\begin{aligned}
x(t+1)& =A_ix(t), \;  x(t_0)=x_0 \\
y(t)& =C_ix(t)+w(t)
\end{aligned}\label{eq:CorruptedAutoSwitchingSystem}
\end{equation}
where $w(t)\in \mathbb{S}^p_\sigma$ represents the $\sigma$-sparse attack on the sensor measurements.

\begin{rem}
	Let two nominal autonomous linear systems $S_i$ and $S_j$ be given (that is, in (\ref{eq:CorruptedAutoSwitchingSystem}) the attack vectors are such that $w_i(t)=w_j(t)=\mathbf{0}$, $\forall \, t\in \mathbb{N}$). When both $S_i$ and $S_j$ have initial condition in the origin (i.e., $x_{0i}=x_{0j}=\mathbf{0}$), the output evolutions are identically zero. Thus, for autonomous systems, the distinguishability between different modes can not be required for any initial state. Therefore, the following distinguishability notion is considered, in which the possibility of both autonomous linear systems having initial condition in the origin, is excluded.
\end{rem}

\begin{defn}
	(\citet{1184923vidal}) Two autonomous linear systems $S_i$ and $S_j$, $(i,j)\in Q \times Q$, are distinguishable, if there exists $\tau \in \mathbb{N}$ such that, for any pair of initial states $x_{0i}$ and $x_{0j}$ (with $x_{0i}\neq 0$ or $x_{0j}\neq 0$), $y_i|_{[0,\tau-1]} \neq y_j|_{[0,\tau-1]}$. The nominal autonomous linear systems $S_i$ and $S_j$ are called indistinguishable if they are not distinguishable.
	\label{def:AutoDist}
\end{defn}

Our aim is to investigate under which conditions it is possible to determine the current mode of the autonomous switching system in (\ref{eq:CorruptedAutoSwitchingSystem}) (without knowing the continuous initial state), when the continuous output signal is corrupted. 
In order to do so, we extend Definition \ref{def:AutoDist} to take into account the presence of the (unknown) attack on sensors.

\begin{defn}
	Two autonomous linear systems $S_i$ and $S_j$, $(i,j)\in Q \times Q$, are $\sigma$-securely distinguishable
	if there exists $\tau \in \mathbb{N}$ such that, for any pair of initial states $x_{0i}$ and $x_{0j}$ (with $x_{0i}\neq 0$ or $x_{0j}\neq 0$), and for any pair of $\sigma$-sparse attack vectors $w_i|_{[0,\tau-1]}\in \mathbb{CS}_{\sigma}^{\tau p}$ and $w_j|_{[0,\tau-1]}\in \mathbb{CS}_{\sigma}^{\tau p}$, $y_i|_{[0,\tau-1]}\neq y_j|_{[0,\tau-1]}$.
\end{defn}

\begin{figure}
	\centering
	\includegraphics[width=0.4\linewidth]{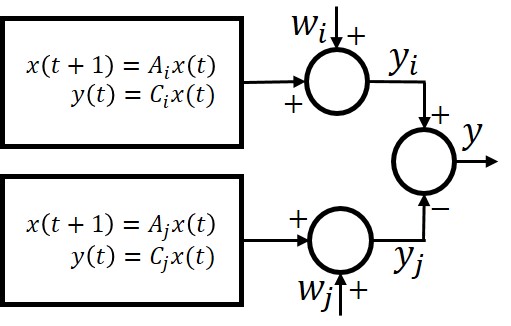}
	\caption{The extended autonomous linear system $S_{ij}$.}
	\label{fig:AutoSystem}
\end{figure}

As already described in the previous section, in order to compare the output evolutions of two autonomous linear systems $S_i$ and $S_j$, $(i,j)\in Q \times Q$, we consider the augmented linear system $S_{ij}$, depicted in Fig. \ref{fig:AutoSystem}, which is fully described by the pair $(A_{ij},C_{ij})$, such that:
\begin{equation}
\begin{aligned}
A_{ij}=\begin{bmatrix}
A_i & \mathbf{0} \\
\mathbf{0}   & A_j
\end{bmatrix}, \; \;  C_{ij}=\begin{bmatrix}
C_i & -C_j
\end{bmatrix}\label{eq:IGmatrix_auto}
\end{aligned}
\end{equation}
with $A_{ij}\in \mathbb{R}^{2n \times 2n}$,  $C_{ij}\in \mathbb{R}^{p \times 2n}$.
The following $2n$ steps-observability matrix is associated with the augmented system $S_{ij}$:
\begin{equation}
\begin{aligned}
O_{ij}& = \begin{bmatrix}
O_i^{(2n-1)} & -O_j^{(2n-1)}
\end{bmatrix} 
\end{aligned}
\end{equation}
where $O_{ij} \in \mathbb{R}^{2np \times 2n}$.

First, we recall the result concerning the distinguishability of two nominal autonomous systems.
\begin{thm}
	(\citet{1184923vidal}) Two nominal autonomous linear systems $S_i$ and $S_j$, $(i,j)\in Q \times Q$, are distinguishable for any initial state $x_{0i}$ and $x_{0j}$  (with $x_{0i}\neq \mathbf{0}$ or $x_{0j}\neq \mathbf{0}$) if and only if $\mathrm{rank}(O_{ij})=2n$.
\end{thm}

\begin{assum}\label{ass:Dist}
	In this section, we assume that, for any pair $(i,j)\in Q \times Q$, the linear systems $S_i$ and $S_j$ are distinguishable.
\end{assum}
\begin{rem}
	In Assumption \ref{ass:Dist} we consider a number of observations $\tau = 2n$ (i.e., equal to the state space's dimension of the augmented system $S_{ij}$). It can be noted that, adding further observations (i.e., beyond $2n$ samples) would not increase the rank of the observability matrix $O_{ij}^{(2n-1)}$ because, due to the Cayley-Hamilton theorem, they would be a linear combination of the first $2n$ components.
\end{rem}

\begin{prop}
	Two autonomous linear systems $S_i$ and $S_j$, $(i,j)\in Q \times Q$,  are $\sigma$-securely distinguishable if and only if, for any set $\Gamma\subseteq \{1,\dots,\gamma\}$ with $|\Gamma|\leq 2\sigma$, the matrix $\overline{O}_{ij, \Gamma}$ obtained from the pairs $(A_i,\overline{C}_{i,\Gamma})$ and $(A_j,\overline{C}_{j,\Gamma})$ has full column rank.\label{prop:secureAuto}
\end{prop}

\begin{pf}
	Assuming to collect $2n$ observations, the output of the augmented linear system $S_{ij}$, as shown in Fig. \ref{fig:AutoSystem}, is:
	\begin{equation}
	\begin{aligned}
	y_i|_{[0, 2n -1]} & - y_j|_{[0, 2n -1]}  = \begin{bmatrix}
	O_i^{(2n-1)} \; \; -O_j^{(2n-1)}
	\end{bmatrix} \begin{bmatrix}
	x_{0i} \\
	x_{0j}
	\end{bmatrix} \\
	& + w_i|_{[0, 2n -1]} - w_j|_{[0, 2n -1]} \\
	& = O_{ij} \begin{bmatrix}
	x_{0i} \\
	x_{0j}
	\end{bmatrix} + w_i|_{[0, 2n -1]} - w_j|_{[0, 2n -1]}
	\end{aligned}\label{eq:CombinedSystem}
	\end{equation}
	Let us rearrange equation (\ref{eq:CombinedSystem})	as:
	\begin{equation}
	\begin{aligned}
	Y_i - Y_j & = O_{ij} \begin{bmatrix}
	x_{0i} \\
	x_{0j}
	\end{bmatrix}
	+ W_i - W_j \\
	& = O_{ij}^{(2n-1)} x_0 + \Delta W
	\end{aligned}
	\end{equation}
	where $W_i=w_i|_{[0, 2n -1]} \in \mathbb{CS}_{\sigma}^{2n p}$, $W_j=w_j|_{[0, 2n -1]} \in \mathbb{CS}_{\sigma}^{2n p}$, $	\Delta W \in \mathbb{CS}_{2\sigma}^{2n p}$,  $Y_i=y_i|_{[0, 2n -1]}$, $Y_j=y_j|_{[0, 2n -1]}$ and $x_0 = \begin{bmatrix}
	x_{0i}^\top \; \; x_{0j}^\top
	\end{bmatrix}^\top$.
	
	\noindent By contradiction, assume that for any set $\Gamma=\{1,\dots,\gamma\}$, $|\Gamma|\leq 2\sigma$, the observability matrix $\overline{O}_{ij,\Gamma}$ obtained from the two pairs $(A_i,\overline{C}_{i,\Gamma})$ and $(A_j,\overline{C}_{j,\Gamma})$ has full column rank (i.e. $\mathrm{ker}(\overline{O}_{ij,\Gamma})=\{\mathbf{0}\}$),
	and that there exist two initial states $x_{0i}$ and $x_{0j}$ ($x_{0i}\neq \mathbf{0}$ or $x_{0j}\neq \mathbf{0}$) and a pair of attack vectors $W_i \in \mathbb{CS}_{\sigma}^{2n p}$, $W_j \in \mathbb{CS}_{\sigma}^{2n p}$,
	such that $Y_i=Y_j$.
	Let $\Gamma=\{k\in \{1,\dots,2n p\} \,| \; \Delta V_{k}  \neq 0\}$ be the set of indexes $k\in \{1,\dots,2n p\}$, for which the corresponding $k$-th element of $\Delta W$ is different from zero (i.e. $\Delta W_{k}  \neq \mathbf{0}$ ), thus $|\Gamma|\leq 2\sigma$.
	$\overline{O}_{ij, \Gamma}x_0=0 \Leftrightarrow x_0 \in \mathrm{ker}(\overline{O}_{ij,\Gamma})$, which is a contradiction.
	
	Assume now that there exists a set of indexes $\Gamma$, $|\Gamma|\leq 2\sigma$, for which the observability matrix $\overline{O}_{ij,\Gamma}$ obtained from the two pairs $(A_i,\overline{C}_{i,\Gamma})$ and $(A_j,\overline{C}_{j,\Gamma})$ has not full column rank (i.e. there exists $x_0 \neq \mathbf{0}$ such that $x_0 \in \mathrm{ker}(\overline{O}_{ij,\Gamma})$), and that there exist two initial states $x_{0i}$ and $x_{0j}$ ($x_{0i}\neq \mathbf{0}$ or $x_{0j}\neq \mathbf{0}$) and a pair of attack vectors $W_i \in \mathbb{CS}_{\sigma}^{2n p}$, $W_j \in \mathbb{CS}_{\sigma}^{2n p}$,
	such that $Y_i\neq Y_j$. $\Gamma=\{k\in \{1,\dots,2n p\} \,| \; \Delta V_{k}  \neq 0\}$ is the set of indexes $k\in \{1,\dots,2n p\}$, for which the corresponding $k$-th element of $\Delta W$ is different from zero (i.e. $\Delta W_{k}  \neq \mathbf{0}$). Assume now to partition the set $\Gamma$ as $\Gamma=\Gamma_i \cup \Gamma_j$ such that  $|\Gamma_i|\leq \sigma$ and $|\Gamma_j|\leq \sigma$. For the sake of clarity consider $O_{ij}x_0=v$ and assume that the pair $(W_i,W_j)$ is such that:
	\begin{equation}
	\begin{aligned}
	W_{i,k} & = \bigg \{ 
	\begin{array}{lr}
	-v_k & \text{if } k\in \Gamma_i \\
	0 & \text{otherwise}
	\end{array}, \\
	W_{j,k} & = \bigg \{ 
	\begin{array}{lr}
	v_k & \text{if } k\in \Gamma_j \\
	0 & \text{otherwise}
	\end{array}
	\end{aligned}
	\end{equation}
	where $W_{i,k}$, $k\in \{1,\dots,2n p\}$, indicates the $k$-th element of the vector $W_i\in \mathbb{CS}_{\sigma}^{2n p}$.
	Thus $Y_i - Y_j=O_{ij}x_0+\Delta W=\mathbf{0}$, which contradicts the initial assumption of distinguishability.	
	\hfill $\square$
\end{pf}


\begin{rem}
	Proposition \ref{prop:secureAuto} corresponds to an observability notion stronger than the classical one, for the augmented system $S_{ij}$. This means that for the linear system $S_{ij}$ the observability property must be satisfied even after the removal of a proper number of output signals (in particular, after the removal of the corrupted sensors).
\end{rem}

Proposition \ref{prop:secureAuto} gives a bound on the cardinality of the set of attacked sensors (that is, for the sparsity of the attack vector).
In particular, it is trivial to check that the following condition has to be satisfied: $2\sigma <  p$.

\begin{rem}
	This condition corresponds to the maximum number of correctable errors (i.e., the maximum number of attackable sensors) derived in \citet{Tabuada6727407} for the discrete-time linear system. Here, we obtain the same bound for secure mode distinguishability of switching systems. 
\end{rem}

\section{Numerical results}\label{sec:NumResult}
In this section, in order to show the applicability of the proposed conditions, we apply them to a DC/DC boost converter. We make use of the hybrid model described in \citet{Sanfelice7070893}, in which the behavior of the DC/DC boost converter is modeled by means of three discrete modes (two corresponding to the open switch, one corresponding to the closed switch), thus $Q=\{1,2,3\}$.
When the switch is open, two dynamical systems $S_1$ and $S_3$ can be active, depending on the diode conducting or not. When the switch is closed, a single dynamics $S_2$ can be considered. The dynamical systems are described by the following matrices:
\begin{equation}
\begin{aligned}
A_1 & ={\footnotesize \begin{bmatrix}
-1/RC & 1/C\\
-1/L & 0
\end{bmatrix}}, \, B_1 ={\footnotesize \begin{bmatrix}
0 \\ 1/L
\end{bmatrix}}  \\
A_2 & ={\footnotesize \begin{bmatrix}
-1/RC &0\\
0 & 0
\end{bmatrix}}, \; B_2=B_1 \\
A_3 & ={\footnotesize \begin{bmatrix}
-1/RC &0\\
0 & 0
\end{bmatrix}}, \; B_3={\footnotesize \begin{bmatrix}
0 \\ 0
\end{bmatrix}} \\
\end{aligned}
\end{equation}

We assume to consider the following output matrices:
\begin{equation}
\begin{aligned}
C_1=C_3={\footnotesize \begin{bmatrix}
0 & 1\\ 0 & 1 \\ -1/RC & 1/C
\end{bmatrix}}, \, C_2={\footnotesize \begin{bmatrix}
1 & 0\\ 1 & 0 \\ -1/RC & 1/C
\end{bmatrix}}
\end{aligned}
\end{equation}

The model provided in \citet{Sanfelice7070893} takes into account continuous-time linear systems. We consider here their discretized version, to model the situation where the sensors send the measurement signals with a time triggered strategy. 
If the systems were autonomous, for any pair $(i,j)\in Q \times Q$, the linear systems $S_i$ and $S_j$ would be distinguishable for any initial state $x_{0i}$ and $x_{0j}$ (with $x_{0i}\neq \mathbf{0}$ or $x_{0j}\neq \mathbf{0}$), as $\mathrm{rank}(O_{ij})=4$, for any $(i,j)\in Q \times Q$.
In this case, if the sensors were corrupted, we should test the condition given in Proposition \ref{prop:secureAuto} for $\sigma $-secure distinguishability. The number of attacked sensors has to satisfy the condition $2\sigma <  p$, therefore the attacker can have access to no more than one sensor or, in other words, (at least) two sensors must be secure.
In order for $S_1$ and $S_2$ to be $1$-securely distinguishable, for any set $\Gamma_1=\{1,2\}$, $\Gamma_2=\{1,3\}$, $\Gamma_3=\{2,3\}$, the matrix $\overline{O}_{12, \Gamma_k}^{(2n-1)}$, $k=1,2,3$, obtained from the pairs $(A_1,\overline{C}_{1,\Gamma_k})$ and $(A_2,\overline{C}_{2,\Gamma_k})$ should be full column rank. 
The results are shown in the first column of Table \ref{tab:ObsRank}.
\begin{table}[h]
	\begin{center}
		\caption{$1$-secure distinguishability of $(S_i,S_j)$}\label{tab:ObsRank}
		\begin{tabular}{cccc}
			$\Gamma$ & $\mathrm{rank}(\overline{O}_{12, \Gamma}^{(2n-1)})$ & $\mathrm{rank}(\overline{O}_{13, \Gamma}^{(2n-1)})$ & $\mathrm{rank}(\overline{O}_{23, \Gamma}^{(2n-1)})$ \\\hline
			$\Gamma_1$ & 4 & 4 & 2\\
			$\Gamma_2$ & 3 & 3 & 2\\
			$\Gamma_3$ & 3 & 3 & 2 \\ \hline
		\end{tabular}
	\end{center}
\end{table}

\vspace*{-0.4cm}
Actually, we can conclude that if $S_1$ and $S_2$ were autonomous, they would not be $1$-securely distinguishable, due to the rank loss for some combinations of sensors. The same holds for the pairs $(S_1,S_3)$ and $(S_2,S_3)$.
However, since the dynamical systems are not autonomous, we can verify the conditions provided in Section \ref{sec:Controlled}.
In particular, assuming that the actuator is secure, we can check if the conditions in Theorem \ref{thm:IFAC:ActuatorSensor} are satisfied.
Both conditions in Theorem  \ref{thm:IFAC:ActuatorSensor} are satisfied, therefore $S_1$ and $S_2$ are $\sigma 0-$securely distinguishable with respect to generic inputs and for all $\sigma-$sparse attacks on sensors, with $\sigma=1$.

\section{Conclusions}
Motivated by the fact that switching systems are an important mathematical formulation for dealing with CPSs, in this paper we investigate under which conditions it is possible to estimate the discrete state of a switching system, when only the continuous output is accessible, and the discrete information is not available. We consider the case in which both sensor measurements and actuator inputs may be compromised by malicious sparse attacks, and we define under which conditions any two discrete modes are securely distinguishable.
The aim of our future work is to propose a computational efficient estimator of the discrete state of switching systems when sensor measurements and/or actuator signals are corrupted by sparse malicious attacks. In addition, a more realistic scenario will be investigated, in which bounded process and measurement noises are also considered in the model of the switching system.

\bibliography{ref}

\end{document}